\documentclass[10pt,aps,prb,twocolumn,superscriptaddress]{revtex4-2}
\usepackage{amsmath}
\usepackage{fixltx2e}
\usepackage{hyperref}
\usepackage{graphicx}
\usepackage{subfigure}
\usepackage{xspace}

\usepackage[dvipsnames]{xcolor}

\usepackage{amsfonts}
\usepackage{mathrsfs}
\usepackage{dsfont}
\usepackage{mathtools}
\usepackage[capitalize]{cleveref}
\usepackage{simplewick}
\usepackage{hyperref}
\usepackage{float}
\usepackage[normalem]{ulem}

\usepackage{tabularx}
\usepackage{amstext} %
\usepackage{array}   %
\usepackage{multirow}

\newcolumntype{L}{>{$}l<{$}} %
\newcolumntype{C}{>{$}c<{$}} %

\begin{document}

\newcommand\kr[1]{\textcolor{blue}{(KR: #1)}}
\newcommand\dk[1]{\textcolor{magenta}{(DK: #1)}}
\newcommand\rv[1]{\textcolor{orange}{(RV: #1)}}
\renewcommand\sb[1]{\textcolor{red}{(SB: #1)}}
\newcommand\ar[1]{\textcolor{brown}{(AR: #1)}}
\newcommand{\corout}[1]{{\color{black}\sout{#1}}}

\title{Combined experimental and theoretical study of hydrostatic (He-gas) pressure effects in $\alpha$-RuCl$_3$}

\author{B. Wolf}
\affiliation{Physikalisches Institut, Goethe-Universit\"at, 60438 Frankfurt (M), Germany}
%\author{Ch. Nabeta}
%\affiliation{Physikalisches Institut, Goethe-Universit\"at, 60438 Frankfurt (M), Germany}
%\affiliation{Leibniz-Institut Festk\"{o}rper- und Werkstoffforschung (IFW) Dresden, 01171 Dresden, Germany}
\author{D. A. S. Kaib}
\affiliation{Institut für Theoretische Physik, Goethe-Universit\"at, 60438 Frankfurt (M), Germany}
\author{A. Razpopov}
\affiliation{Institut für Theoretische Physik, Goethe-Universit\"at, 60438 Frankfurt (M), Germany}
\author{S. Biswas}
\affiliation{Institut für Theoretische Physik, Goethe-Universit\"at, 60438 Frankfurt (M), Germany}
\author{K. Riedl}
\affiliation{Institut für Theoretische Physik, Goethe-Universit\"at, 60438 Frankfurt (M), Germany}
\author{S. M. Winter}
\affiliation{Institut für Theoretische Physik, Goethe-Universit\"at, 60438 Frankfurt (M), Germany}
\affiliation{Department of Physics and Center for Functional Materials, Wake Forest University, North Carolina 27109, USA}
\author{R. Valent\'{\i}}
\affiliation{Institut für Theoretische Physik, Goethe-Universit\"at, 60438 Frankfurt (M), Germany}
\author{Y. Saito}
\affiliation{Physikalisches Institut, Goethe-Universit\"at, 60438 Frankfurt (M), Germany}
\author{S. Hartmann}
\affiliation{Physikalisches Institut, Goethe-Universit\"at, 60438 Frankfurt (M), Germany}
\author{E. Vinokurova}
\affiliation{Institut f\"ur Festk\"orper- und Materialphysik, Technische Universit\"at Dresden, 01062 Dresden, Germany,}
\affiliation{Institut für Festkörperforschung, Leibniz IFW-Dresden, 01069 Dresden, Germany}
\author{T. Doert}
\affiliation{Faculty of Chemistry and Food Chemistry, Technische Universit\"at Dresden, 01062 Dresden, Germany}
\author{A. Isaeva}
\affiliation{Institut für Festkörperforschung, Leibniz IFW-Dresden, 01069 Dresden, Germany}
\affiliation{Van der Waals-Zeeman Institute, Department of Physics and Astronomy, University of Amsterdam, Science Park 094, 1098 XH Amsterdam, The Netherlands}
\author{G. Bastien}
\affiliation{Institut für Festkörperforschung, Leibniz IFW-Dresden, 01069 Dresden, Germany}
\affiliation{Charles University, Faculty of Mathematics and Physics, Department of Condensed Matter Physics, Ke Karlovu 5, 121 16, Prague 2, Czech Republic}
\author{A. U. B. Wolter}
\affiliation{Institut für Festkörperforschung, Leibniz IFW-Dresden, 01069 Dresden, Germany}
\author{B. B\"{u}chner}
\affiliation{Institut für Festkörperforschung, Leibniz IFW-Dresden, 01069 Dresden, Germany}
\affiliation{Institut für Festk\"orper- und Materialphysik and W\"urzburg-Dresden Cluster of Excellence ct.qmat, Technische Universit\"at Dresden, 01062 Dresden, Germany}
\author{M. Lang}
\affiliation{Physikalisches Institut, Goethe-Universit\"at, 60438 Frankfurt (M), Germany}

\date{\today}

\begin{abstract}
We report a detailed experimental and theoretical study on the effect of hydrostatic pressure on various structural and magnetic aspects of the layered honeycomb antiferromagent $\alpha$-RuCl$_{3}$. This material has been intensively studied recently with the objective of probing fundamental aspects of Kitaev physics. Through measurements of the magnetic susceptibility, $\chi$, performed under almost ideal hydrostatic-pressure conditions by using helium as a pressure-transmitting medium, we find that the phase transition to zigzag-type antiferromagnetic order at $T_N$ = 7.3\,K can be rapidly suppressed to about 6.1\,K at a weak pressure of about 94\,MPa. A further suppression of $T_N$ with increasing pressure is impeded, however, due to the occurrence of a pressure-induced structural transition at $p \geq$ 104\,MPa, accompanied by a strong dimerization of Ru-Ru bonds, which gives rise to a collapse of the magnetic susceptibility. Whereas the dimerization transition is strongly first order, as reflected by large discontinuous changes in $\chi$ and pronounced hysteresis effects, the magnetic transition under varying pressure and magnetic field also reveals indications for a weakly first-order transition. We assign this observation to a strong magnetoelastic coupling in this system. Measurements of $\chi$ under varying pressure in the paramagnetic regime ($T > T_N$) and before dimerization ($p <$ 100\,MPa) reveal a considerable increase of $\chi$ with pressure. These experimental observations are consistent with the results of \textit{ab-initio} Density Functional Theory (DFT) calculations on the pressure-dependent structure of $\alpha$-RuCl$_{3}$ and the corresponding pressure-dependent magnetic model. This model includes the nearest-neighbor Kitaev ($K$) and Heisenberg ($J$) couplings as well as the off-diagonal anisotropic exchanges $\Gamma$ and $\Gamma '$. We find a pressure-induced strengthening of $|\Gamma|$ and $|J|$ and a simultaneous weakening of $|K|$ and $|\Gamma '|$, leading to a destabilization of the magnetic order.
%The results of our combined experimental and theoretical investigations spoil the initial hope for accessing a Kitaev spin liquid phase by the suppression of magnetic order via hydrostatic pressure.
Comparative susceptibility measurements on a second crystal showing two consecutive magnetic transitions instead of one, indicating the influence of stacking faults, reveal that by the application of different temperature-pressure protocols the effect of these stacking faults can be temporarily overcome, transforming the magnetic state from a multiple-$T_N$ into a single-$T_N$ state.
\end{abstract}


\maketitle

%-------------------%
%  1. Introduction and motivation %
%-------------------%
\section{Introduction}
The layered honeycomb material $\alpha$-RuCl$_{3}$ is one of the prime candidates to probe fundamental aspects of Kitaev physics~\cite{Kitaev2006anyons,Jackeli2009}. This assessment has motivated numerous studies on this and other materials despite the occurrence of magnetic order at low temperatures, reflecting additional magnetic interactions beyond the pure Kitaev coupling \cite{chaloupka2013zigzag,rau2014generic,winter2016challenges}. There have been various attempts to access the range where Kitaev physics prevails in $\alpha$-RuCl$_{3}$ by suppressing the magnetic order via variation of
temperature  \cite{Sandilands15, Nasu16, do2017majorana, Widmann19} or  application
of a magnetic field \cite{johnson2015monoclinic, yadav2016kitaev, Sears17, wolter2017field, baek2017evidence, wang2017magnetic, Zheng17, Ponomaryov17, banerjee2018excitations, hentrich2018UnusualPhononHeat, kasahara2018majorana, Janssen19, kaib2019dynamicalresponse}.
Specially under strong scrutiny has been the region beyond magnetic order under application of an in-plane magnetic field,
where the possibility of an intermediate field-induced spin liquid phase is being
controversially discussed \cite{baek2017evidence,wang2017magnetic,banerjee2018excitations,kasahara2018majorana, kaib2019dynamicalresponse, hentrich2019LargeThermalHall,czajka2021OscillationsThermalConductivitya, kasahara2018unusual, balz2019finite,ponomaryov2020NatureMagnetic, bachus2020thermodynamic,tanaka2020ThermodynamicEvidenceFieldangle,bachus2021AngledependentThermodynamics, yamashita2020sample, bruin2021RobustnessThermalHall}.
As most striking measurements are thermal Hall conductivity measurements \cite{kasahara2018majorana,ponomaryov2020NatureMagnetic, bachus2020thermodynamic,tanaka2020ThermodynamicEvidenceFieldangle,bachus2021AngledependentThermodynamics, yamashita2020sample, bruin2021RobustnessThermalHall} where some studies reported half-integer quantization \cite{kasahara2018majorana,bachus2021AngledependentThermodynamics, yamashita2020sample, bruin2021RobustnessThermalHall} as predicted  for the Kitaev honeycomb model under a magnetic field \cite{Kitaev2006anyons}.

An aspect that has lately gained increased attention is the interplay of Kitaev interactions and magnetoelastic coupling \cite{metavitsiadis2020phonon,ye2020phonon,kaib2021magnetoelastic}, which is expected to influence the thermal Hall conductivity measurements \cite{vinkler2018approximately,ye2018quantization,ye2021PhononHallViscosity}.
In $\alpha$-RuCl$_{3}$, large anisotropic magnetostriction effects \cite{gass2020field,schonemann2020thermal,kocsis2022} and a high sensitivity of phonon properties to magnetism ~\cite{hentrich2018UnusualPhononHeat,hentrich2020high,li2021giant} were recently observed. These imply the presence of a substantial magnetoelastic coupling and a strong dependence of the various exchanges on modifications in the lattice structure \cite{kaib2021magnetoelastic}.

The application of external pressure
is in this context an interesting avenue, both to further investigate the magnetoelastic coupling itself, as well as to tune the material away from magnetic order.
In fact, pressure has proven an excellent tuning parameter for exploring exotic quantum states in a variety of materials, with organic charge-transfer salts \cite{Gati2016breakdown, Kurosaki05,  Shimizu16} being among the most prominent examples.
The pressure studies performed on $\alpha$-RuCl$_{3}$ until now \cite{Cui17, biesner2018detuning, bastien2018pressure, wang2018pressure} have revealed that the magnetic ground state becomes unstable under pressure giving way to a non-magnetic phase at $p \geq p_c$ with different critical pressure values reported of $p_c$ $\sim$ 0.2\,GPa \cite{bastien2018pressure} or $\sim$ 1\,GPa \cite{biesner2018detuning}. This high-pressure phase has been assigned to a state with triclinic symmetry \cite{bastien2018pressure} characterized by a strong dimerization of Ru-Ru bonds \cite{bastien2018pressure, biesner2018detuning}. Less clear is, however, the response of the magnetic state to weak and moderate pressures $p \leq p_c$. Whereas measurements of the specific heat, magnetization and NMR \cite{Cui17, wang2018pressure}, all of which were performed by using oil as a pressure-transmitting medium, revealed a slight increase of $T_N$ with pressure, thermal expansion measurements, carried out at ambient pressure, predict a rapid suppression of $T_N$ with pressure and a complete suppression of magnetic order around 0.3\,GPa \cite{He18}.
To address the question of whether or not magnetic order can be reduced or even fully suppressed by the application of pressure, the peculiarities of this material have to be taken into account. $\alpha$-RuCl$_{3}$ has a layered structure with weakly van-der-Waals-coupled honeycomb layers, which implies a strong tendency to various types of structural transitions -- common to the whole family of metal trihalides \cite{McGuire17} -- and the formation of stacking faults \cite{johnson2015monoclinic,kim16, cao2016low}. Therefore, deviations from truly hydrostatic pressure conditions can be an issue. Such problems %\dk{Isn't this too aggressive, saying the others had problems?}
can be minimized by using a pressure-transmitting medium with a solidification temperature as low as possible in order to reduce the thermal expansion mismatch between the frozen pressure medium and the sample upon cooling.

Here we present a thorough magnetic susceptibility study on $\alpha$-RuCl$_{3}$ single crystals under almost ideal hydrostatic pressure conditions by using helium as a pressure-transmitting medium. Our observations are supported by \textit{ab-initio} Density Functional Theory (DFT) calculations on the pressure dependence of the structure and the so-derived pressure-dependent magnetic model, which we find to reproduce consistently the evolution of the experimental susceptibility. The salient results of our study can be summarized as follows: (i) The magnetic ordering temperature becomes rapidly reduced with pressure from $T_N$ = 7.3\,K at ambient pressure to about 6.1\,K at $p \approx$ 94\,MPa but cannot be fully suppressed to $T_N$ = 0 on further increasing the pressure due to the occurrence of the pressure-induced dimerization transition \cite{bastien2018pressure, biesner2018detuning} at $T_d \approx$ 45\,K (upon cooling) for $p \approx$ 110\,MPa. (ii) Thorough investigations of the magnetic transition at $T_N$ at varying pressures and magnetic fields reveal clear indications for a weak first-order transition. (iii) The magnetic susceptibility in the paramagnetic regime ($T > T_N$) and before dimerization ($p <$ 100\,MPa) considerably increases with pressure. Based on our model calculations, we assign this effect to a pressure-induced strengthening of the nearest-neighbor Heisenberg $|J|$ and off-diagonal anisotropic $\Gamma$ coupling and a simultaneous weakening of the Kitaev $|K|$ and anisotropic $\Gamma '$ couplings.
(iv) By applying different temperature-pressure protocols, the magnetic state of a single crystal, showing two consecutive magnetic transitions, can be transformed into a single-$T_N$ state.

The paper is organized as follows. Section II presents the methods used for the experimental (Sec.\,IIA) and theoretical (Sec.\,IIB) investigations. Experimental results are given in Sec.\,III with the sample characterization in Sec.\,IIIA, the pressure and field dependence of the magnetic transition in Sec.\,IIIB, the pressure-induced dimerization transition in Sec.\,IIIC, the pressure-temperature phase diagram in Sec.\,IIID, and the effect of pressure on a multiple-$T_N$ state in Sec.\,IIIE. In Sec.\,IV we present our theoretical calculations and
in Sec.\,V we discuss the results of the experimental and theoretical studies by focussing on the structural phase transition at $T_s$ in Sec.\,VA, the pressure dependence of the magnetic susceptibility in the paramagnetic regime in Sec.\,VB, the phase transition into zigzag-type magnetic order in Sec.\,VC, the structural transition at $T_d$ in Sec.\,VD, and the healing effect of pressure-temperature treatments on a multiple-$T_N$ state in Sec.\,VE. Finally, the conclusions are summarized in Sec.\,VI.

\section{Methods}
\subsection{Experiments}
Single crystals of $\alpha$-RuCl$_{3}$ were grown using the chemical transport reactions method \cite{hentrich2018UnusualPhononHeat}.
Measurements were performed on two crystals (\#S50, \#SS-Ru-1-CVT, labeled from here on as \#1 and \#2, respectively) prepared by following different protocols, see Appendix Sec.\,A. These crystals differ by the occurrence of a single magnetic phase transition (\#1) as opposed to two magnetic transitions (\#2), indicating the presence (or higher concentration) of stacking faults for crystal \#2 \cite{cao2016low}. The susceptibility was measured by using a commercial superconducting quantum interference device (SQUID) magnetometer (MPMS, Quantum Design) equipped with a CuBe pressure cell (Unipress Equipment Division, Institute of High Pressure Physics, Polish Academy of Science). The pressure cell is connected via a CuBe capillary to a room temperature $^{4}$He-gas compressor, serving as a gas reservoir, which enables temperature sweeps to be performed at $p \approx$ const. conditions. The use of helium as a pressure-transmitting medium ensures truly hydrostatic pressure conditions over the wide range of temperatures and pressures where helium is in its liquid phase. Even when entering the solid phase upon crossing the solidification line $T_{sol}$($p$) (with $T_{sol}$(100\,MPa) = 13.6\,K for example, see black dotted line in Fig.\,4 below), which is accompanied by a pressure drop of about 5\% $\pm$ 0.2\% for 20\,MPa $\leq \textit{p} \leq$ 350\,MPa \cite{Grilly59}, the deviations from hydrostatic conditions are very small. This is due to the low solidification temperature of helium, implying a small thermal expansion mismatch between sample and frozen pressure medium upon cooling, and the small shear modulus of solid helium \cite{Haziot13}. The susceptibility data were corrected for the contribution of the sample holder, including the CuBe pressure cell determined independently. The single crystals were also characterized by low-temperature specific heat measurements performed by using a heat-pulse relaxation method in a Physical Properties Measurement System (PPMS, Quantum Design). In order to obtain the specific heat of $\alpha$-RuCl$_{3}$, the temperature-dependent addenda was subtracted from the measured specific-heat values in the sample measurements.

\subsection{Theory}

We perform full structural optimization under hydrostatic pressure by means of density functional theory (DFT) calculations, as implemented in VASP~\cite{PhysRevB.47.558}. The exchange-correlation functional is approximated by generalized gradient approximation (GGA)~\cite{PhysRevLett.77.3865}, and the  correlation corrections are included through the Dudarev~\cite{Dudarev} scheme with the effective Coulomb repulsion $U_{\rm eff} = 1.5$~eV.  A planewave cut-off of 650\,eV is used for the expansion of the basis functions along with a 10$\times$10$\times$10 \textbf{k}-mesh for the Brillouin zone sampling. The spin-orbit coupling (SOC) is taken into account by performing fully relativistic calculations.	Additionally, we include van der Waals (vdW) corrections through the DFT+D2 method of Grimme \cite{Grimme}. A force convergence criteria of $10^{-3}$ eV/\AA\ is used for all-three components of the forces for all the atoms in the unit cell. The initial guess structure for the structural optimizations is	the experimentally confirmed monoclinic structure (space group $C2/m$) determined by X-ray diffraction of single crystals~\cite{cao2016low}. Relaxations are performed assuming the zigzag antiferromagnetic (zzAFM) configuration.
Note that the calculated pressure range is far wider than that covered in the experiment due to the underbinding problem of the GGA+SOC methods as was observed in other Kitaev candidate materials, e.g, in iridates and rhodates \cite{hermannPRB2018,hermannPRB2019} and due to negligible changes in calculated lattice geometries in the experimental pressure range ($\sim 0.1$ \AA).
Though these shortcomings result in a larger transition pressure to dimerization than in the experimental observations, the structural trends obtained this way are able to reproduce consistently the experimental trends of susceptiblity and N\'{e}el temperature, as we will show below.

%	\subsubsection{Pressure-dependent  $j_\text{eff}$ Hamiltonian}
	
	For each relaxed structure, we extracted then an effective spin $j_{\rm eff}=1/2$ Hamiltonian up to third nearest neighbor sites via the ``projED'' method \cite{winter2016challenges,riedl2019abinitio}.
	First, we determined the hopping parameters between the $4d$ ruthenium atoms via wannierization of non-relativistic band structure calculations with the Full Potential Local Orbital \cite{fplo} code (FPLO) within the GGA approximation \cite{PhysRevLett.77.3865}. These parameters allow then to construct an electronic Hubbard Hamiltonian $\mathcal{H}_{\rm tot}=\mathcal{H}_{\rm hop}+\mathcal{H}_{\rm SOC} + \mathcal{H}_{\rm int}$, where $\mathcal{H}_{\rm hop}$ contains the structure-specific \textit{ab-initio} single-particle hopping parameters. The spin-orbit coupling effects in $\mathcal{H}_{\rm SOC}=\lambda \sum_i \mathbf{L}_i \cdot \tilde{\mathbf S}_i$ (where $\tilde{\mathbf S}_i$ denotes the electron spin) were treated in the atomic approximation with a spin-orbit coupling strength $\lambda=0.15\,$eV \cite{montalti2006handbook}. For the two-particle interaction $\mathcal{H}_{\rm int}$ the employed Coulomb repulsion $U_{t_{2g}} = F_0 + (4/49)(F_2 + F_4) = 1.68\,$eV (where $F_k$ are the radial Slater integrals) and Hund's coupling $J_{t_{2g}} = (3/49)F_2 + (20/441)F_4 = 0.29\,$eV follow constrained RPA results \cite{kaib2022electronic, eichstaedt2019deriving} for the experimental $C2/m$ structure of $\alpha$-RuCl$_{3}$.
In a second step, we then solved the two-site, five-orbital Hubbard Hamiltonian by exact diagonalization and the effective spin Hamiltonian extracted via projection onto the low-energy subspace, i.e.\ $\mathcal{H}_{\rm eff} = \mathbb{P} \mathcal{H}_{\rm tot}\mathbb{P} = \sum_{i\mu j \nu} J_{ij}^{\mu \nu} S_i^\mu S_j^\nu$ with $\mu$, $\nu \in \{x,y,z\}$ and $j_{\text{eff}}=1/2$ operators $\mathbf S_i$.
This procedure with the same model parameters $\lambda$, $U_{t_{2g}}$ and $J_{t_{2g}}$ was successfully employed previously by some of the authors to calculate magnetic exchange parameters for $\alpha$-RuCl$_{3}$ structures under uniaxial strain \cite{kaib2021magnetoelastic}.

Finally, to compare the so-derived pressure-dependent spin models to our experiments, we employ exact diagonalization methods. For finite-temperature calculations of the magnetic susceptibility, we took advantage of the recently developed orthogonalized finite-temperature Lanczos method  (OFTLM) \cite{OFTLM}.
This method combines the random sampling of the finite-temperature Lanczos method \cite{FTLM} (FTLM) with numerically exact results for few lowest-energy states to improve the accuracy at low temperatures. This was necessary for reliably computing the magnetization $M$, since no conservation of $M$ could be utilized in the present models, leading to worse convergence of standard FTLM than in more symmetric models \cite{schnack2020accuracy}.
For all presented results, we employed $N_v=7$ exact low-energy states, $M=45$ Krylov steps for the FTLM sum, and at least $R\ge 300$
random initial states. Statistical errors due to a finite $R$ were estimated by a jackknife resampling procedure. The calculations were performed on a 24-site honeycomb cluster with full point group symmetry. We note that finite-size effects prohibit true symmetry breaking in our simulations, and only hints of phase transitions for the thermodynamic limit can be obtained. However, we will show that the experimentally observed pressure-induced changes can be qualitatively and consistently reproduced.

\section{Experimental results}
\subsection{Sample characterization}

\begin{figure}[h]
    %\centering
	\includegraphics[width=1.0\columnwidth]{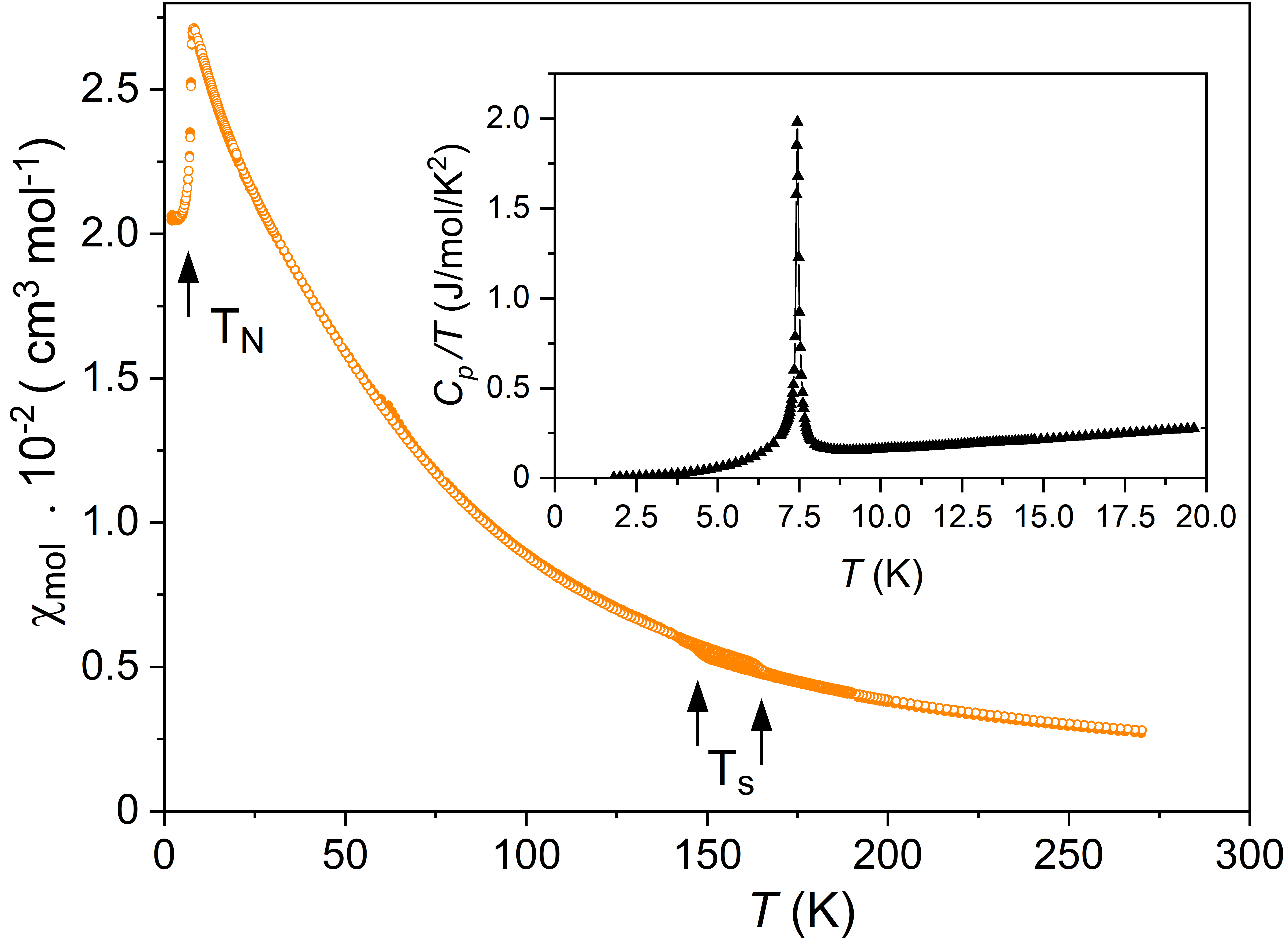}
	\caption{Molar magnetic susceptibility of $\alpha$-RuCl$_3$ (crystal \#1) as a function of temperature at an in-plane field $B$ = 0.1\,T. The full (open) orange circles represent data points taken with decreasing (increasing) temperature. The inset exhibits specific heat data, plotted as $C_p$/$T$ vs.\,$T$ for the same crystal.}
	\label{Fig1}
\end{figure}

 Figure 1 shows data of the magnetic susceptibility taken at ambient pressure (labelled $p$ = 0 from here on) on crystal \#1. The results essentially reproduce published data \cite{do2017majorana, bastien2018pressure} but excel by revealing a single, sharp drop in $\chi$ at $T_N$ = 7.33\,K, signaling the high quality of this crystal. This assessment is also corroborated by low-temperature specific heat data on the same crystal, shown in the inset of Fig.\,1, yielding a very sharp phase transition anomaly at $T_N$. The susceptibility data in Fig.\,1 disclose a second, distinctly smaller anomaly centered around 157\,K with a pronounced hysteresis of width $\sim$ 17\,K upon cooling and warming (see arrows near $T_s$ in Fig.\,1). We assign this feature to the first-order structural transition at $T_s$ from a high-temperature monoclinc $C2/m$ structure \cite{johnson2015monoclinic, cao2016low} to a low-temperature structure, the symmetry of which is still under debate, see Sec.\,VA. \\

\subsection{Pressure- and field dependence of the magnetic transition}

\begin{figure}[t]
    %\centering
	\includegraphics[width=0.8\columnwidth]{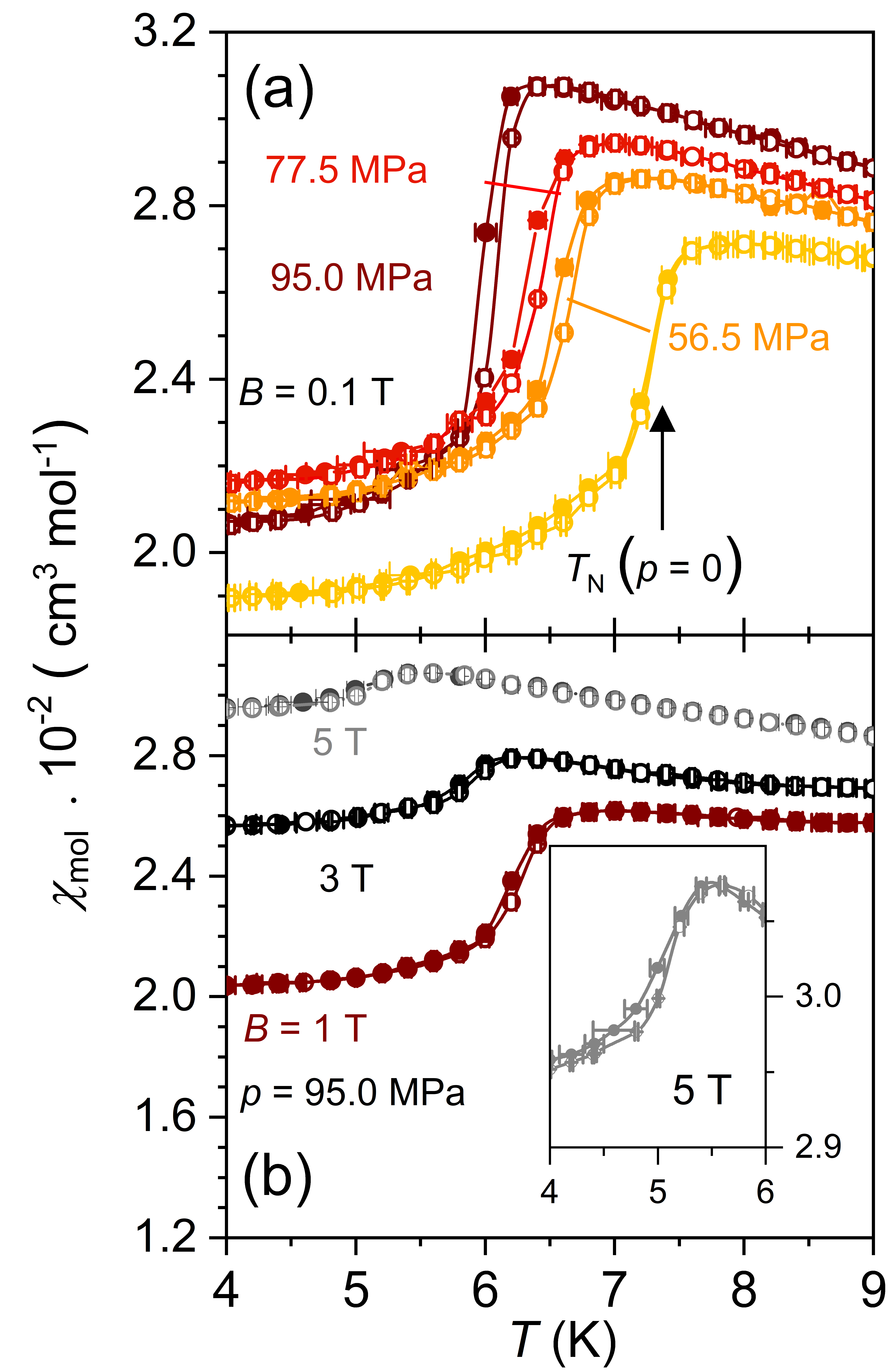}
	\caption{Magnetic susceptibility of $\alpha$-RuCl$_3$ (crystal \#1). Upper part (a) shows data as a function of temperature at an in-plane field $B$ = 0.1\,T for various pressure values: $p$ = 0 (dark yellow circles), 56.5\,MPa (orange circles), 77.5\,MPa (red circles), and 95.0\,MPa (dark brown circles). The full (open) symbols represent the data taken with decreasing (increasing) temperature. Lower part (b) shows data as a function of temperature at $p$ = 95.0\,MPa for various magnetic fields applied parallel to the planes: $B$ = 1\,T (brown circles), 3\,T (black circles) and 5\,T (grey circles). The inset exhibits the data taken at 5\,T on enlarged scales.}
	\label{Fig2}
\end{figure}

Figure 2 shows the low-temperature part of the magnetic susceptibility on expanded scales for crystal \#1 measured at different pressure values at an in-plane field $B$ = 0.1\,T (Fig.\,2a) and at different magnetic fields of $B$ = 1, 3, and 5\,T at $p$ = 95.0\,MPa (Fig.\,2b). Note that for all gas pressures applied here ($p \leq$ 95.0\,MPa), $T_{sol}$($p$) of the pressure-transmitting medium helium lies above 9.5\,K (see Fig.\,4 below) to avoid any interference of the solidification process with the magnetic transition in $\alpha$-RuCl$_{3}$.

\begin{figure}[h]
    %\centering
	\includegraphics[width=1.0\columnwidth]{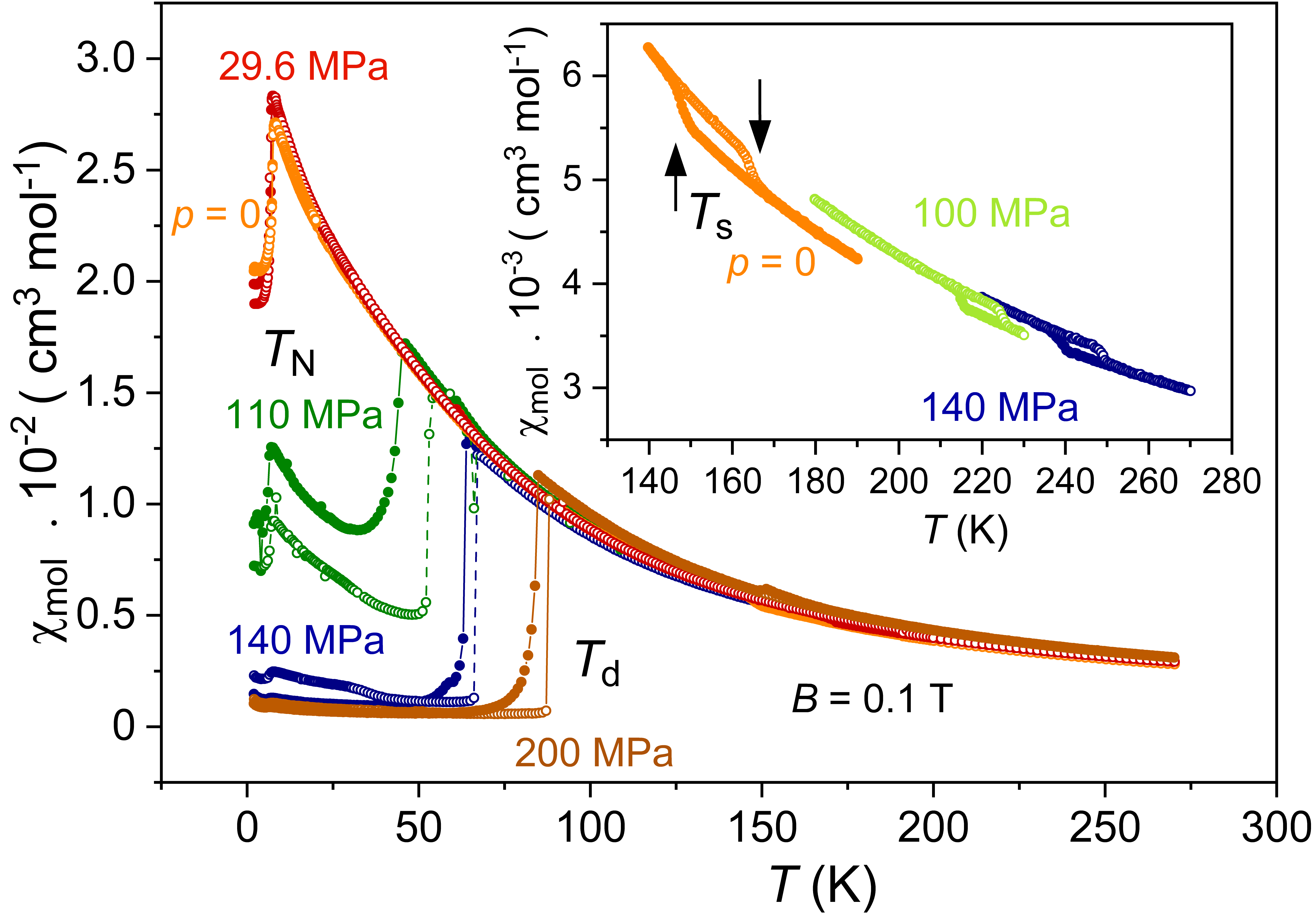}
	\caption{Magnetic susceptibility of $\alpha$-RuCl$_3$ (crystal \#1) for temperatures 2\,K $\leq T \leq$ 270\,K at an in-plane field $B$ = 0.1\,T for various pressure values: $p$ = 0 (orange circles), 29.6\,MPa (dark red circles), 110\,MPa (dark green circles), 140\,MPa (dark blue circles) and 200\,MPa (brown circles). The full (open) symbols represent the data taken with decreasing (increasing) temperature. The inset shows a blow-up of the data around $T_s$ taken at ambient pressure (orange circles), at 100\,MPa (light green circles) and at 140\,MPa (dark blue circles).}
	\label{Fig3}
\end{figure}

At each pressure value data have been taken upon decreasing and increasing temperature. Several observations can be made: (1) Whereas the data at $p$ = 0 lack any clear indication of hysteretic behavior, the phase transition anomaly revealed at finite pressure of 56.5, 77.5 and 95\,MPa \footnote{These pressure values correspond to the pressure in the solid phase of helium. According to Ref.\,\onlinecite{Grilly59} the solidification is accompanied by a pressure drop of about 5\% $\pm$ 0.2\% for the pressure range 20\,MPa $\leq \textit{p} \leq$ 350\,MPa.} all show a small but distinct hysteresis upon cooling and warming. Hysteretic behavior is revealed also on increasing the field to $B$ = 1\,T, 3\,T and 5\,T at $p$ = 95\,MPa, see the inset of Fig.\,2b for the data at 5\,T on enlarged scales. (2) With increasing pressure, the susceptibility at $T \geq T_N$ increases considerably in a monotonic manner. This growth in $\chi$ in the paramagnetic regime is accompanied by an increase in the size of the drop of $\chi$($T$) upon cooling through $T_N$. (3) The data in Fig.\,2a reveal a clear suppression of $T_N$ with increasing pressure. By identifying $T_N$ with the maximum in d($\chi \cdot T)$/d$T$ (see Sec.\,B in the Appendix), we find $T_N$ = 7.33\,K ($p$ = 0), 6.65\,K (56.4\,MPa), 6.5\,K (77.5\,MPa), and 6.04\,K (95.0\,MPa). \\

\begin{figure}[h]
    %\centering
	\includegraphics[width=0.8\columnwidth]{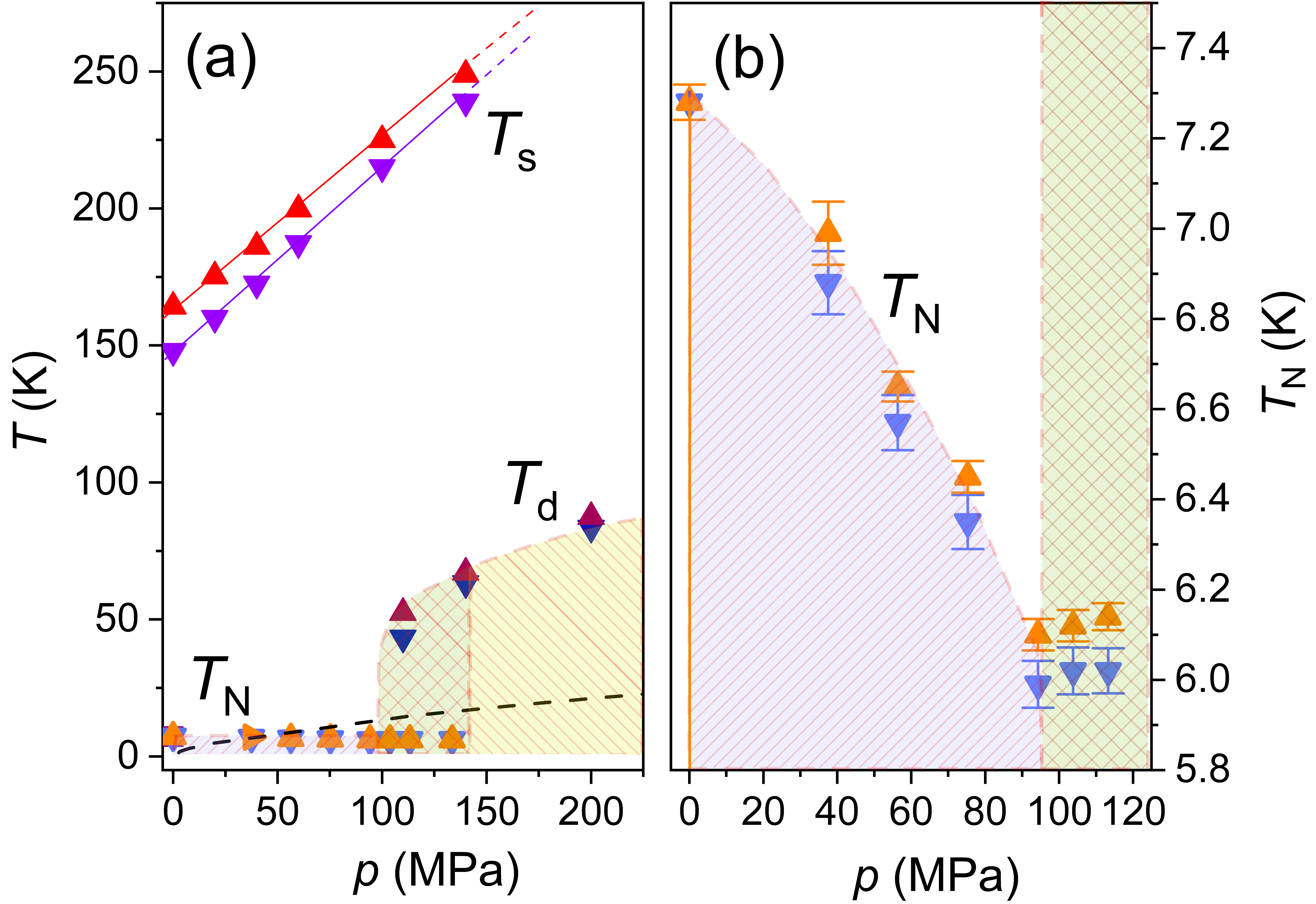}
	\caption{(a) $\textit{p-T}$ phase diagram of $\alpha$-RuCl$_3$ (crystal \#1) including a high-temperature structural transition at $T_s$, the transition to zigzag-type antiferromagnetic order below $T_N$ and the structural transition characterized by the dimerization of Ru-Ru bonds below $T_d$. Whereas the transitions at $T_s$ and $T_d$ are strongly first order, the transition at $T_N$ shows indications for a weak first-order character. The cross-hatched green area at intermediate pressure 95\,MPa $\leq p \leq$ 140\,MPa marks a range of inhomogeneous phase coexistence. The black broken line in (a) represents the solidification line $T_{sol}$($p$) of the pressure-transmitting medium helium. (b) Blow-up of the variation of $T_N$ with pressure for $p \leq$ 125\,MPa. Up (down) triangles in (a) and (b) indicate data points collected upon increasing (decreasing) temperature.}
	\label{Fig4}
\end{figure}

\subsection{Pressure-induced dimerization and collapse of magnetic susceptibility}
Figure 3 shows the susceptibility for crystal \#1 at various pressure values 0 $\leq p \leq$ 200\,MPa over the whole temperature range investigated. At $p$ = 29.6\,MPa, the data reveal essentially the same behavior as at $p$ = 0 except a slightly enhanced $\chi$($T$) at low temperatures and a reduced $T_N$. On increasing the pressure to 110\,MPa, 140\,MPa, and 200\,MPa, however, a sharp drop in $\chi$($T$) occurs at intermediate temperatures which is accompanied by a pronounced hysteresis upon cooling and warming. We assign this drop to the first-order structural transition at $T_d$ characterized by the dimerization of Ru-Ru bonds \cite{bastien2018pressure, biesner2018detuning} which is accompanied by the collapse of the magnetic susceptibility. For the cooling runs, we obtain $T_d$ = 43.28\,K (110\,MPa), 63.17\,K (140\,MPa) and 83.35\,K (200\,MPa). On further cooling the susceptibility data at $p$ = 110\,MPa and 140\,MPa show some remnants of the non-dimerized phase, i.e., an enhanced susceptibility which grows upon cooling followed by a drop at the magnetic transition temperature $T_N$. For $p$ = 110\,MPa, $\chi$($T$) drops at $T_d$ to about half its value at $T\geq T_d$; a similar reduction occurs for the jump at $T_N$ as compared to that at $p$ = 0, cf.\,Fig.\,2. In contrast, for $p$ = 140\,MPa, the reduction of both the susceptibility for $T < T_d$ and the jump at $T_N$ amounts to about 90\%. %We assign these observations for $p$ = 110\,MPa and 140\,MPa to an inhomogeneous coexistence for $T \leq T_d$ of a non-magnetic dimerized phase and a non-dimerized magnetic phase.
The inset of Fig.\,3 shows a blow-up of the data around the structural transition at $T_s$ for the various pressures. The data reveal an extraordinarily strong increase of $T_s$ with pressure which is accompanied by a significant reduction of the width of the hysteresis.\\

\subsection{Pressure-temperature phase diagram}
In Fig.\,4 we compile the data for $T_s$, $T_N$ and $T_d$ of crystal \#1 in a temperature-pressure phase diagram. The diagram discloses several remarkable features. In the pressure range investigated, the structural phase transition at $T_s$ follows, to a very good approximation, a linear increase with pressure at an extraordinarily high rate of d$T_s$/d$p$ = (614 $\pm$ 10)\,K/GPa. Such a strong increase of $T_s$ is consistent with the relatively large volume expansion at this transition and small entropy change \cite{Widmann19, He18,Park16}. This increase in $T_s$ is accompanied by a uniform reduction in the width of the hysteresis with pressure indicating a complete closure around 350\,K. In contrast to this $p$-linear variation of $T_s$, there is a strikingly discontinuous evolution of the phases associated with $T_N$ and $T_d$ at lower temperatures. For small pressures $p <$ 95\,MPa, the system orders antiferromagnetically with $T_N$ becoming rapidly suppressed with pressure, cf.\,Fig.\,4b for a blow-up of the $T_N(p)$ data. For the initial slope we find (d$T_N$/d$p$)$_{p\rightarrow 0}$ = -(10.9 $\pm$ 0.5)\,K/GPa. This rate becomes progressively stronger with increasing pressure reaching a value of about (d$T_N$/d$p$) = -(25 $\pm$ 3)\,K/GPa around $p$ = 95.0\,MPa.

Upon further increasing the pressure from 95\,MPa to 200\,MPa the material's ground state changes drastically by showing a progressive collapse of the susceptibility due to the formation of Ru-Ru dimers \cite{bastien2018pressure, biesner2018detuning} for $T < T_d$ accompanying a structural transition into a triclinic ($P\bar 1$) low-temperature phase \cite{bastien2018pressure}. This transition is strongly first order as reflected by the sharp discontinuity in $\chi$($T$) and the pronounced hysteresis upon cooling and warming. At intermediate pressures 95\,MPa $\leq p \leq$ 140\,MPa, the data indicate an inhomogeneous phase coexistence below $T_d$ of the dimerized, non-magnetic phase with the non-dimerized, magnetic phase (cross-hatched area in Fig.\,4). The transition temperature shows a strong increase with pressure, starting at $T_d$ = 43.5\,K (measured upon cooling) at 100\,MPa and reaching a value of $T_d \approx$ 83.3\,K at 200\,MPa. The width of the hysteresis at $T_d$ amounts to 9.4\,K at 100\,MPa and reduces to 4.0\,K at 200\,MPa.\\

\subsection{Effect of pressure on a multiple-$T_N$ state}

\begin{figure}[h]
    %\centering
	\includegraphics[width=1.0\columnwidth]{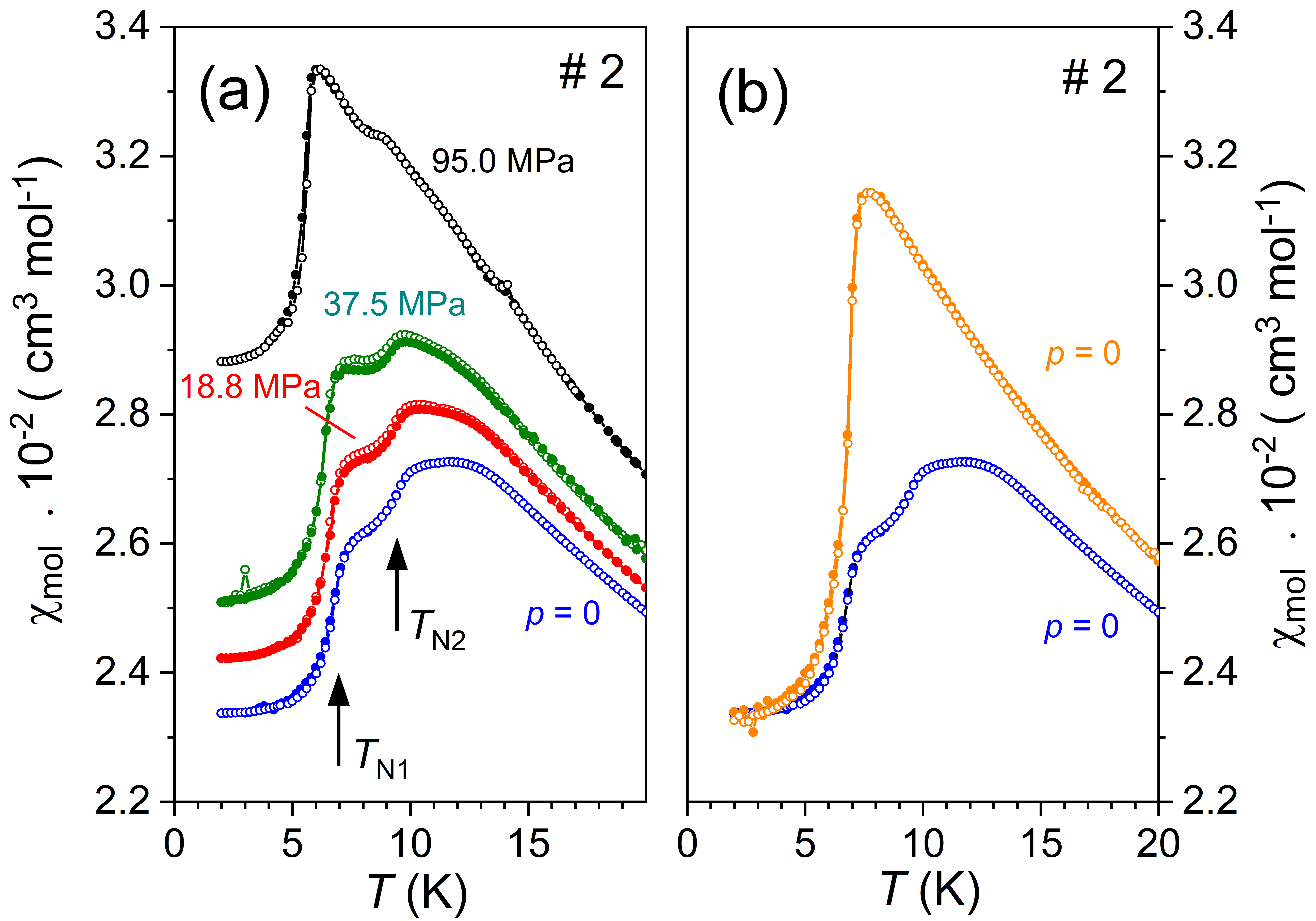}
	\caption{(a) Magnetic susceptibility of $\alpha$-RuCl$_3$ crystal \#2 at $p$ = 0 (blue circles) measured during the initial cooling procedure from room temperature down to 2\,K and in the subsequent heating run from 2\,K up to 20\,K. For the subsequent runs at finite pressure of 18.8\,MPa (red circles), 37.6\,MPa (dark green circles) and 95\,MPa (black circles), the pressure was applied at 20\,K and data were taken upon cooling (full symbols) and warming (open symbols). (b) Magnetic susceptibility of crystal \#2 at $p$ = 0 after following different protocols: The blue circles represent the data taken after the initial cool down whereas the orange circles correspond to data obtained after warming the sample to 50\,K and applying a pressure of 140\,MPa for a limited period of time of approximately 24\,hours before releasing the pressure to $p$ = 0, see text for details.}
	\label{Fig5}
\end{figure}

The $\chi$($T,p$) measurements on crystal \#1 discussed above were supplemented by corresponding experiments on crystal \#2 which shows two magnetic transitions, cf.\,Fig.\,5a. The figure displays the results of a set of consecutive measurements on this crystal as a function of temperature taken at different pressures 0 $\leq p \leq$ 95\,MPa. The pressure changes between the runs were conducted at 20\,K. The blue symbols represent data taken at $p$ = 0 during the initial cooling procedure and the subsequent heating run. In this 'virgin' state the susceptibility reveals two magnetic transitions of similar size at $T_{N2} \approx$ 9.5\,K and $T_{N1} \approx$ 6.8\,K. With increasing pressure the size of the anomaly at $T_{N1}$ increases while that of $T_{N2}$ decreases. At 95\,MPa, the maximum pressure applied in this set of experiments, the anomaly at $T_{N2}$ is almost completely suppressed. Figure\,5b compares the low-temperature susceptibility of crystal \#2 in its 'virgin' state (blue circles in Fig.\,5(a) and 5(b) with another data set, also taken at ambient pressure (orange circles), obtained after applying the following protocol: The sample was heated up to 50\,K and then slowly pressurized to 140\,MPa. The pressure and temperature were kept constant for approximately 24\,hours. Subsequently, the pressure was released to $p$ = 0 prior to cooling the sample to the base temperature of 2\,K. Data were taken upon cooling (full symbols) and subsequent heating (open symbols) to 20\,K. Figure 5 clearly demonstrates that following this particular temperature-pressure treatment, the magnetic state of crystal \#2 can be drastically changed, now showing a single phase-transition anomaly at $T_N$ = 6.9\,K. The size of the transition is very similar to that revealed for crystal \#1, cf.\,Fig.\,2. We stress that this transformation into a single-$T_N$ state is reversible, as the system adopts its original multiple-$T_N$ state, once the crystal had been warmed up to room temperature (not shown).

\section{Theoretical results}

We performed extensive \textit{ab-initio} calculations combined
with exact diagonalization methods on finite clusters of the resulting low-energy magnetic models to qualitatively understand the microscopic behavior of the structure and magnetic interactions under hydrostatic pressure.

\begin{figure}
	  \centering
      \includegraphics[width=1.0\columnwidth]{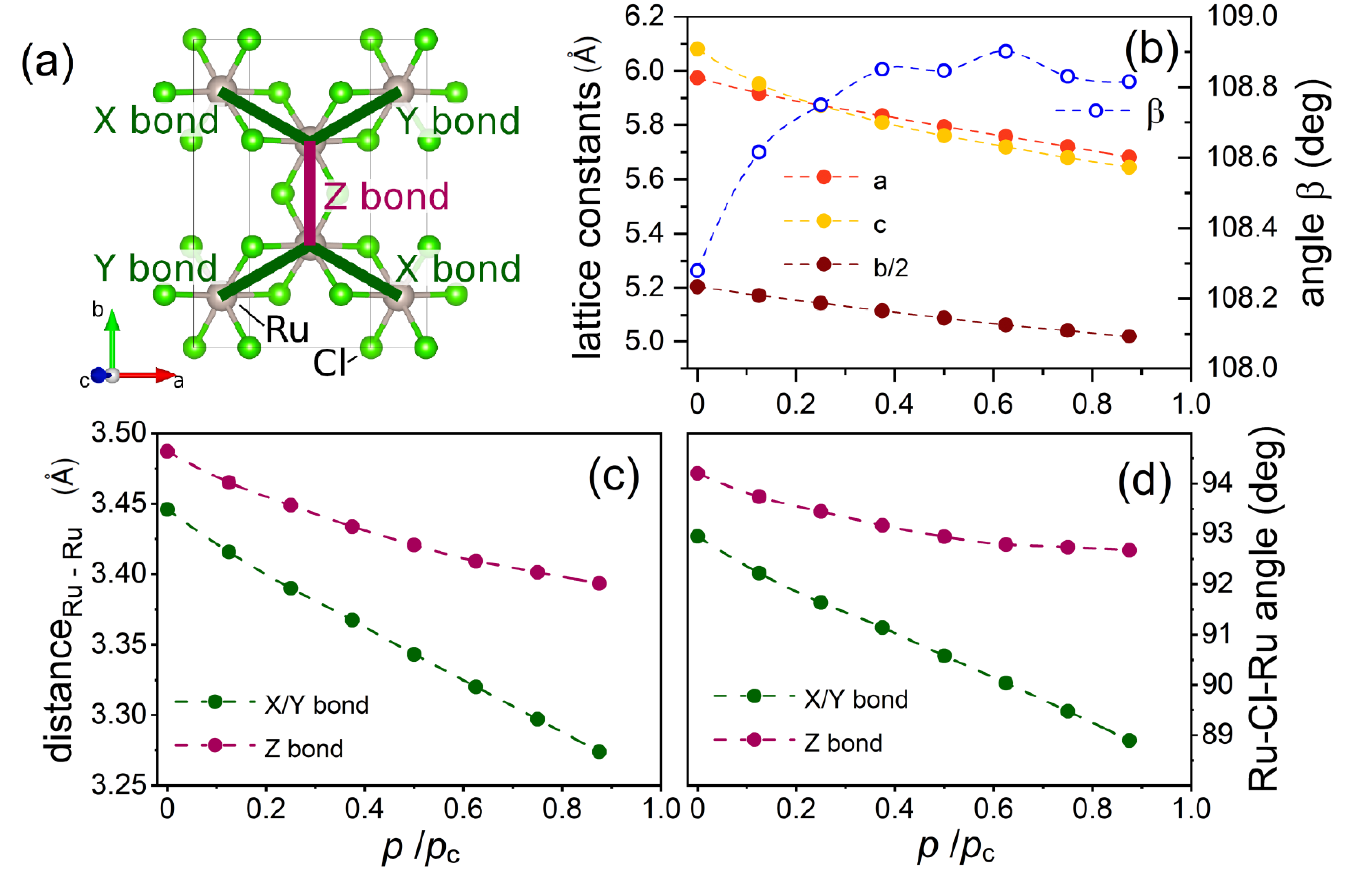}
      \caption{
      (a) Crystal structure of $\alpha$-RuCl$_{3}$ in the C2/m space group. %Green (gray) spheres represent Cl (Ru).
      Bond names X, Y and Z are depicted.
      (b-d) Calculated (DFT+SOC+U) lattice parameters as a function of pressure. (b) Lattice constants $a$, $b$, $c$ (left y-axis) and the monoclinic angle $\beta$ (right y-axis). (c) Ru-Ru interatomic distances for X/Y bonds and Z bonds. (d) Ru-Cl-Ru bond angle around the Z-bond and X(Y)- bond. $p_c$ denotes the critical pressure at which the structure is found to dimerize.}\label{fig:crystal_structure_and_relaxation}
\end{figure}

\subsection{Crystal structure under applied hydrostatic pressure}

The results of the crystal structure relaxation under hydrostatic pressure are summarized in Fig.~\ref{fig:crystal_structure_and_relaxation}.
Before dimerization ($p/p_c<1$, where $p_c$ denotes the critical pressure at which the structure
dimerizes), the crystal structure remains in the $C2/m$ symmetry with zigzag AFM ordering being energetically favored.
The lattice parameters (\cref{fig:crystal_structure_and_relaxation}(b)), as well as Ru-Ru interatomic distances
(\cref{fig:crystal_structure_and_relaxation}(c)) and Ru-Cl-Ru bond angles (\cref{fig:crystal_structure_and_relaxation}(d)), decrease with increasing pressure, while the monoclinic angle $\beta$ between the $a$ and $c$ axes enhances slightly for higher pressures.  The bond ratio Z-bond/X(Y)-bond,
changes from 0.988
at ambient pressure to 0.965 at $p/p_c=0.875$, while
remaining in the  $C2/m$ space group, as shown in Fig.~\ref{fig:crystal_structure_and_relaxation}(a).

A structural phase transition to the dimerized phase
(Z-bond/X(Y)-bond $\approx 1.25$) occurs above $p_c=8$ GPa and the symmetry of the crystal lattice changes from $C2/m$ to $P\bar{1}$ (not shown here). In the dimerized triclinic lattice ($P\bar{1}$), there is only one short bond (either X- or Y-bond) and two large bonds (Y- and Z-bonds, or X- and Z-bonds). We note that while the absolute
value of the critical pressure is overestimated with respect to the
experimental observation for the reasons mentioned in the
Method section, the trends predicted by the calculations are in agreement with experiments as shown below. To emphasize the qualitative agreement with experiment, we show relative pressures $p/p_c$ in Fig.~\ref{fig:crystal_structure_and_relaxation}.

\subsection{Pressure-dependent $j_\text{eff}$ Hamiltonian}
The exchange matrix on the Z-bond (see Fig.~\ref{fig:crystal_structure_and_relaxation}) in the conventionally used cubic $x,y,z$ axes is parametrized as
	\begin{align}
    J_{ij} = \left(\begin{array}{ccc} J & \Gamma & \Gamma^\prime \\ \Gamma & J & \Gamma^\prime \\ \Gamma^\prime & \Gamma^\prime & J+K \end{array} \right).
    \label{eq:mag_parameters}
    \end{align}
    Here $K$ ($J$) is the Kitaev (Heisenberg) coupling, and $\Gamma$, $\Gamma'$ are off-diagonal anisotropic exchanges.
    The X- and Y-bond definitions follow by cyclic permutation of $(x,y,z)$ axes. Note that there are symmetry-allowed additional correction terms $\xi$ and $\zeta$ (see definition in e.g.\ Ref. ~\onlinecite{winter2016challenges}), which are however small at ambient and low pressure, and vanish upon $C_3$-symmetrization.

\begin{figure}
	    \centering
	    \includegraphics[width=\linewidth]{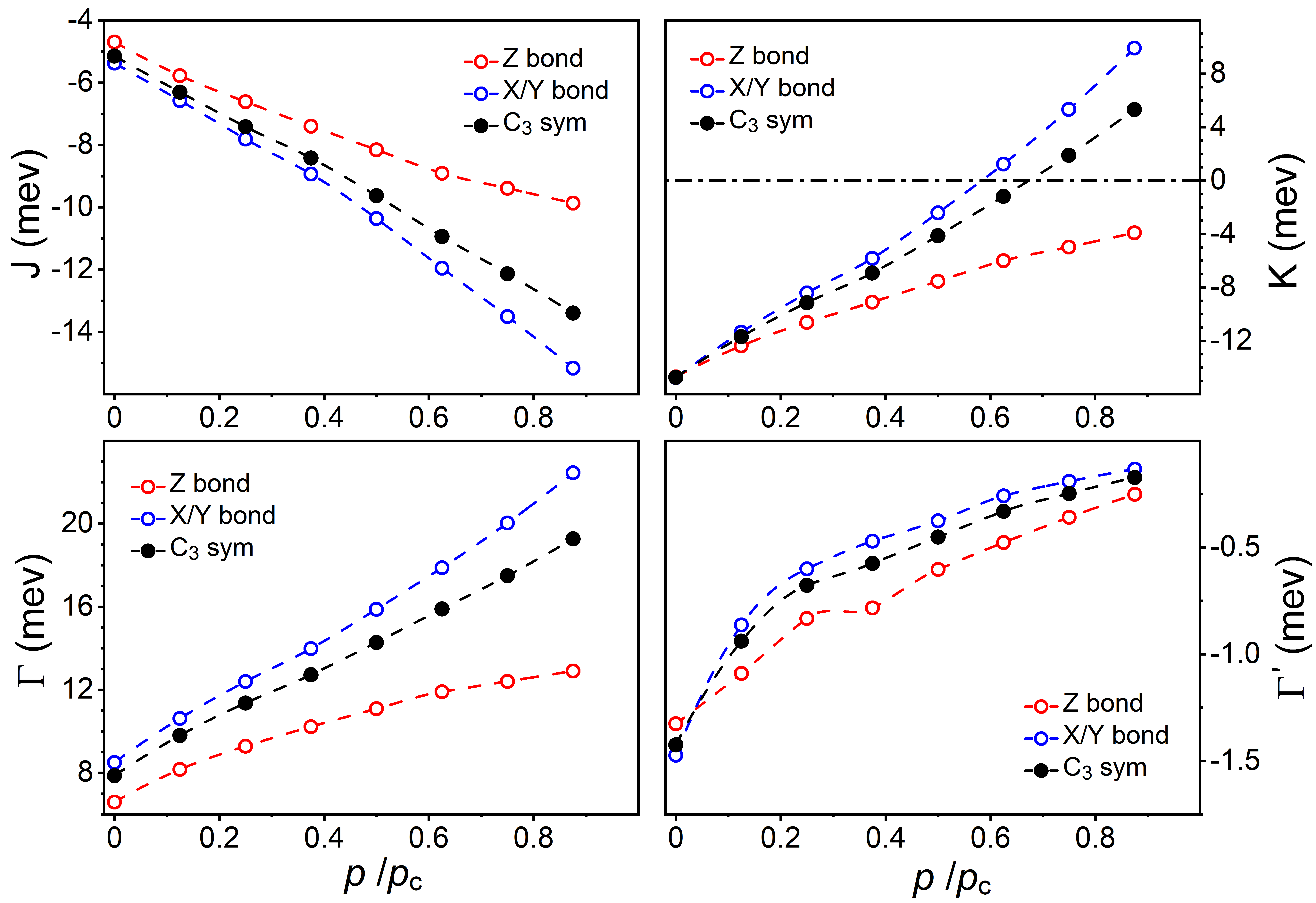}
	    \caption{Nearest-neighbor exchange parameters as a function of pressure obtained with the projED method. % The $C_3$-symmetrized parameters are illustrated in black, while the parameters on a Z- and X-bond are illustrated with red and blue circles respectively, parametrized as in \cref{eq:mag_parameters}.
	    The couplings (parametrized as in \cref{eq:mag_parameters}) on Z- and X/Y-bonds are shown as red and blue circles, respectively, while the $C_3$-symmetrized parameters are illustrated in black.
	    }
	    \label{fig:exchange}
	\end{figure}

    Our pressure-dependent results for the nearest-neighbor exchanges are illustrated in \cref{fig:exchange}, with open points denoting X/Y- and Z-bond results, and filled points the $C_3$-averaged results, where X/Y- and Z-bonds have same coupling strengths.
    %\kr{In Appendix XX we discuss the results for up to third nearest neighbours at ambient and finite pressure. } \dk{see next comment} \kr{Okay. But what do we write in the appendix then other then the table? Or would you give the table in the main text?}
    At ambient pressure, we obtain a model with a dominant ferromagnetic Kitaev ($K<0$) interaction, and smaller $\Gamma>0$, $J<0$ and $\Gamma'<0$ exchanges. This is generally consistent with the established literature on $\alpha$-RuCl$_{3}$ \cite{maksimov2020,laurell2020DynamicalThermalMagnetic}. Through computing spin-spin correlations $\langle \mathbf S_{-\mathbf Q} \cdot \mathbf S_{\mathbf Q} \rangle$ within exact diagonalization, we find that zigzag antiferromagnetic order is correctly reproduced by the obtained $j_\text{eff}=1/2$ models at low temperatures and low pressures.

    Note that the Kitaev interaction $K$ gets a strong positive contribution with increasing pressure, leading to a sign change (i.e.\ antiferromagnetic interaction) for $p\ge 0.6 \,p_c$. In these high-pressure regimes the dominant interactions therefore are Heisenberg $J$ and off-diagonal $\Gamma$. For instance, at $p=0.75 \,p_c$, we have dominant $J\approx -12.1$ and  $\Gamma\approx 17.5$ meV, whereas $K\approx 1.9$, $\Gamma' \approx -0.25$, $\xi\approx -1.5$, $\zeta\approx 0.9$ meV. $\xi$ and $\zeta$ -\,usually ignored in the analysis of Kitaev materials\,- become similarly important as the Kitaev interaction at high pressures near dimerization.

 Longer-range interactions are found to be significantly smaller, both at zero and at finite pressure. We find $g$ values of in-plane $g_{ab}\approx2.3$ and out-of-plane $g_{c^*}\approx 1.9$ at ambient pressure. Their pressure-dependence is found to be rather weak compared to that of the exchange constants, with maximal relative variations (along the scanned pressure range) below $5\%$.

In contrast to the case of compressive uniaxial strain along $c^\ast$, \cite{kaib2021magnetoelastic} for compressive \emph{hydrostatic} pressure the $a$ and $b$ crystal axes are not expanding. For the dominant nearest-neighbor exchange parameters this results in some opposite trends for the individual interactions. Here, the magnitudes of the Kitaev interaction $K$ and off-diagonal $\Gamma^\prime$ decrease, while the Heisenberg interaction $J$ and off-diagonal $\Gamma$ are enhanced in magnitude. Taking into account the signs of the respective coupling, this implies a destabilization of zigzag order, mainly due to the positive contribution to $\Gamma'$. Since the Kitaev coupling $|K|$ decreases with pressure, this destabilization may not necessarily be in favor of a Kitaev spin liquid phase, and instead accompanies a stronger competition with ferromagnetism, due to the strong negative contribution to~$J$.

\section{Discussion} %\textendash\
Below we discuss the results of our experimental and theoretical studies on the various structural and magnetic properties of $\alpha$-RuCl$_{3}$ and the variation of these properties upon the application of hydrostatic pressure.

\subsection{Structural phase transition at $T_s$} We start the discussion by first focussing on the structural transition at $T_s$. A structural phase transition has been observed for many metal trihalide compounds with partially-filled $d$ shells \cite{McGuire17}. For $\alpha$-RuCl$_3$ this transition is from a high-temperature monoclinic $C2/m$ structure \cite{johnson2015monoclinic, cao2016low} to a low-temperature phase the symmetry of which is still under debate. Whereas some reports are indicating a monoclinic $C$2/$m$ low-$T$ structure \cite{johnson2015monoclinic, cao2016low} others are in favor of a rhombohedral $R\overline{3}$ \cite{Park16, Nagai20} structure. According to our susceptibility measurements on crystal \#1 (Figs.\,1 and 3) and thermal expansion measurements at ambient pressure on crystal \#2 (see Sec.\,D in the Appendix), yielding discontinuous length changes accompanied by pronounced hysteresis, this transition is strongly first order. The transition temperature $T_s$ is somewhat lower for crystal \#2, where it is centred around 149\,K, as compared to 157\,K for crystal \#1. In addition, the width of the hysteresis of about 46\,K for \#2 is larger than 17\,K revealed for \#1, cf.\,Fig.\,1. We assign the differences in the transition temperature and the width of the thermal hysteresis to the presence of stacking faults, the concentration of which is presumably higher for crystal \#2 as compared to \#1, see the discussion below on the magnetic properties. By following the evolution of $T_s$ with increasing pressure via magnetic susceptibility measurements on \#1 we find a strictly linear pressure dependence of $T_s$ at an extraordinarily high rate of d$T_s$/d$p$ = (614 $\pm$ 10)K/GPa. This linear evolution of $T_s$ with pressure contrasts with the strongly non-linear variation with pressure of the magnetic- and dimerization transitions at $T_N$ and $T_d$, respectively. This indicates that the structural transition at $T_s$ has only little effect on the broken-symmetry state evolving at low temperatures. \\

\subsection{Pressure dependence of magnetic susceptibility}
Measurements of the magnetic susceptibility under varying pressure in the paramagnetic regime ($T>T_N$) and before dimerization, i.e., $p < p_c \approx$ 100\,MPa, reveal a considerable, monotonous increase of $\chi$ with pressure. This is accompanied by an increasing suppression of $T_N$ and a non-monotonic evolution of $\chi$ with pressure below $T_N$. In what follows, we compare these experimental findings with calculations for the susceptibility based on the model presented above and discuss the implications with respect to the relevant couplings.

For the paramagnetic region $T>T_N$, we can estimate the expected effect of a pressure-induced change of each coupling onto $\chi$ from a high-temperature expansion or equivalently from the dependence of the Weiss constant on these couplings \cite{li2021modified}. These imply that,  for in-plane magnetic fields, $\chi$ can be increased by negative contributions to $J$ and $K$, and by positive contributions to $\Gamma$ and $\Gamma'$.
Hence, from the calculated pressure-dependence of the respective couplings (\cref{fig:exchange}), the trends revealed for $J$, $\Gamma$, $\Gamma'$ all lead to the experimentally observed increase in the high-temperature susceptibility, although the influence of $K$ competes.

\begin{figure}
\centering
\includegraphics[width=0.85\columnwidth]{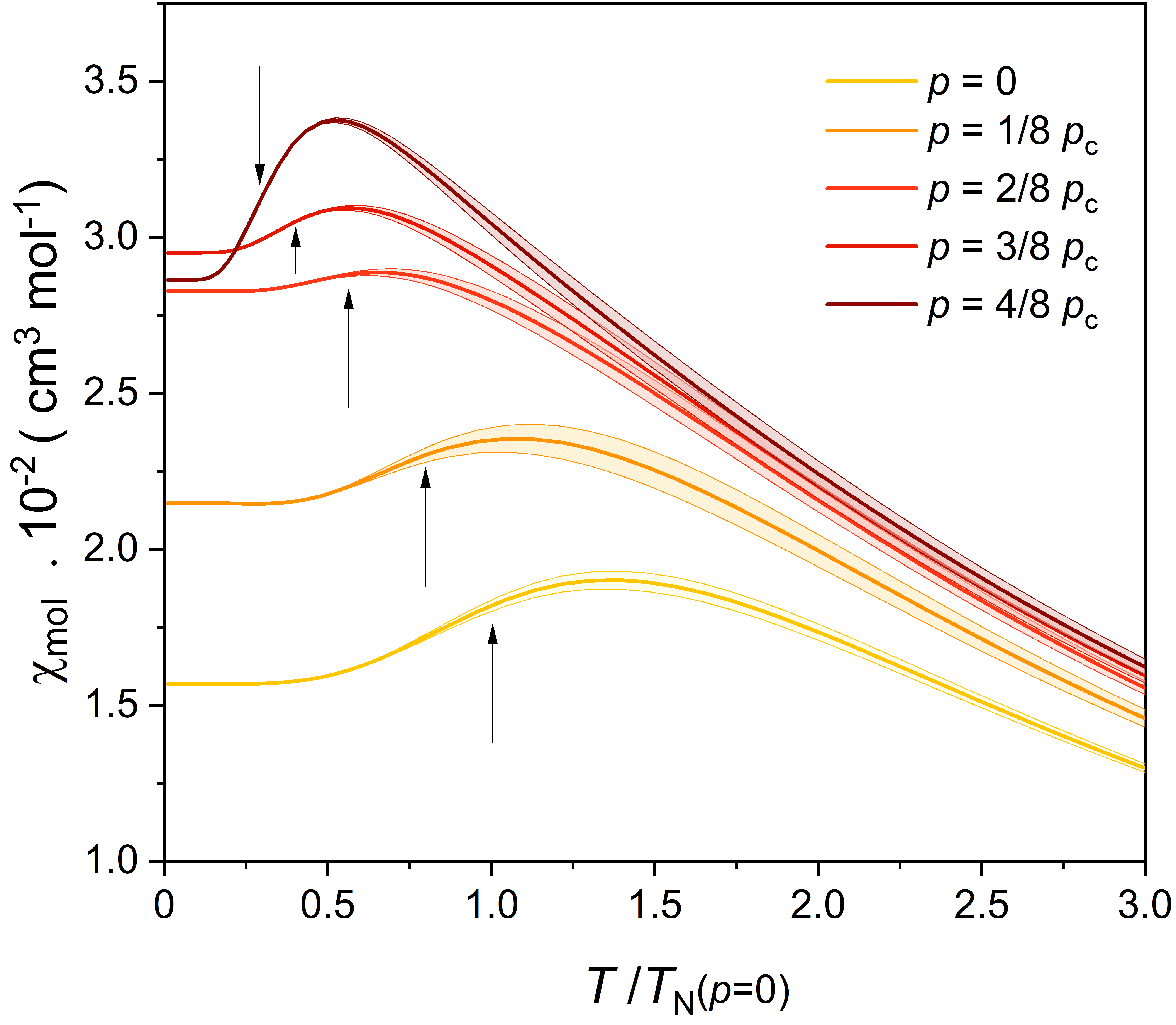}
\caption{Magnetic susceptibilities of the \textit{ab-initio} derived $j_\text{eff}=1/2$ models at different pressures normalized to $p_c$, the critical pressure above which the dimerization transition occurs. The data are computed via OFTLM (see main text). The temperature axis has been normalized to the N\'{e}el temperature at ambient pressure. The arrow for each pressure indicates the Néel temperature, determined by a maximum in $\mathrm d(\chi\cdot T)/\mathrm dT$. Shaded areas indicate estimates of statistical errors ($\pm 1 \sigma$).}
\label{fig:theory_sucept}
\end{figure}

To more accurately also model intermediate and low temperatures, we employ OFTLM as described in the methods section.
We employ the nearest-neighbor $C_3$-symmetrized models discussed above, and a pressure-independent in-plane $g$-value of $g_{ab}=2.3$, as the pressure-dependence of the $g$ tensor was found to be minor compared to that of the exchange couplings.
	In \cref{fig:theory_sucept} we show the temperature-dependent susceptibility $\chi=M/B$ for models at various pressures and  small magnetic field $B=1$\,T (with $\mathbf B\parallel b$).
	Statistical error estimates are shown as shaded areas. The Néel temperature is determined by a maximum in $\mathrm d(\chi\cdot T)/\mathrm dT$, as done in our measurements.
	While the results indicate an overestimation of the absolute Néel temperature $T_N\approx 24$\,K at zero pressure% compared to the measured $T_N\approx 7$\,K
	, we find the qualitative trends to be well in line with our main experimental observations (Figs.~2(a), 3).
	Specifically, with increasing pressure, $T_N$ is monotonically suppressed, while the high-temperature susceptibility ($T\gtrsim T_N$) grows with pressure. The susceptibility at low temperatures $T<T_N$, on the other hand, first grows with pressure but eventually decreases at intermediate pressure (see $p/p_c=\frac{4}{8}$ in \cref{fig:theory_sucept}), which is also consistent with the experiment (cf.~Fig.2(a)).
	
Overall we conclude good agreement with the experiment. The observed dependence of the magnetic susceptibility is consistent with a pressure-induced strengthening of the off-diagonal $\Gamma$ and Heisenberg exchanges and simultaneous weakening of the Kitaev and $\Gamma'$ couplings, leading to destabilization of zigzag magnetic order.

\subsection{Phase transition into zigzag-type antiferromagnetic order at $T_N$} For weak and moderate pressures $p <$ 95\,MPa the system orders antiferromagnetically with an ordering temperature $T_N$ which becomes rapidly suppressed at a rate which increases with pressure reaching a value of about (d$T_N$/d$p$) = -(25 $\pm$ 3)\,K/GPa around $p$ = 95\,MPa. A further suppression of $T_N$ is impeded by the occurrence of the dimerization transition at $T_d$ for $p \geq$ 95\,MPa giving rise to a finite range in the $p-T$ phase diagram of inhomogeneous phase coexistence. An important aspect revealed in the present study relates to the order of the phase transition at $T_N$. The jump-like changes observed in the magnetic susceptibility in $T$-dependent measurements at finite pressure $p \leq$ 95\,MPa and varying magnetic fields $B \leq$ 5\,T, together with the hysteretic behavior detected upon cooling and warming, are clear indications for the first-order character of the transition. In addition, a discontinuous change in $\chi$($T$) at $T_N$ characterizes also the data at $p$ = 0 albeit these data lack clear indications for hysteretic behavior. In this respect we like to mention the results of thermal expansion measurements at ambient pressure also yielding step-like changes in $\Delta L$/$L$ at $T_N$, indicating a first-order phase transition. The appearance of the $\Delta L$/$L$ anomaly, however, depends on the crystal investigated and the measurement technique applied \cite{Widmann19, gass2020field, schonemann2020thermal}. The sharp step-like changes, presented in \cite{schonemann2020thermal} using a fiber Bragg grating method, were observed for crystals with a small amount of stacking faults similar to \#1 in this work.

First-order magnetic transitions are relatively rare -- the vast majority of the known magnetic transitions are of second order. However, there has been a certain number of materials identified where the transition is first order. Two classes can be distinguished. Most common are magnetostructural transitions involving a simultaneous change in the crystal symmetry. Less frequent are magnetoelastic transitions that leave the crystal symmetry unperturbed, (see, e.g., \cite{Guillou18} and references cited therein).

For $\alpha$-RuCl$_3$, it is tempting to assign the first-order nature of the magnetic phase transition to the strong magnetoelastic coupling in this system. This is reflected in the simultaneous occurrence of anomalies in the lattice and magnetic degrees of freedom at the various phase transitions. A strong magnetoelastic coupling is also supported by our model calculations.  Alternatively, a first-order magnetic transition resulting from the action of a finite biquadratic term in the exchange interaction, in addition to the dominant bilinear term, has been discussed \cite{Frazer1965, Rodbell1963}. However, such a scenario seems unlikely for $\alpha$-RuCl$_3$ as it rather applies to highly symmetric systems such as UO$_2$ and MnO \cite{Frazer1965}. In this context, it is relevant that the magnetic spacegroup associated with zigzag magnetic order is generally not a maximal subgroup of the spacegroup of the paramagnetic phase, which is the typical situation for continuous transitions \cite{cracknell2016magnetism}. For example, we can discuss isolated layers with 3-fold rotational and inversion symmetries with space group $P\bar{3}1^\prime$. The zigzag phase is then the magnetic spacegroup $P_S\bar{1}$, with e.g. a doubled unit cell along the $a$-axis. It can be shown that $P\bar{3}1^\prime \supset P\bar{1}1^\prime \supset P_S\bar{1}$ \cite{perez2015symmetry}, which implies a two-step ordering process where rotational symmetry is first broken on decreasing temperature (giving a nematic-like phase), then time-reversal symmetry is finally broken giving zigzag magnetic order. Indeed, this series of transitions was observed in the classical simulations of the Heisenberg-Kitaev models in Ref. \cite{price2013finite}. Under conditions where these two symmetries are broken simultaneously, Ref. \cite{price2013finite} found the single transition becomes first order.  Thus, we conclude that a first-order transition is compatible with the observed magnetic order.

It has been found that some characteristics of first-order magnetic phase transitions are sensitive to compositional adjustments, and that a partial replacement of atoms on the non-magnetic site can result in an almost complete disappearance of the hysteresis \cite{Guillou14}. These observations, along with our findings on the influence hydrostatic pressure has on the effects of stacking faults in $\alpha$-RuCl$_{3}$, suggest the following scenario for crystal \#1: in this crystal there is a small but finite concentration of stacking faults which makes the hysteresis accompanying the weakly first-order transition at $T_N$ unmeasurably small at ambient pressure. Above a certain threshold pressure, however, the system adopts a metastable state rendering the stacking faults ineffective and the hysteresis becomes clearly visible. An immediate consequence of the (weak) first-order character of the magnetic transition, which retains its the first-order character also in fields up to at least 5\,T (cf.\,Fig.\,2(b), is that fluctuations are strongly suppressed raising questions about the relevance of quantum critical fluctuations in such samples as \#1 around the field of $\mu{_0}H_{c1}$ = 7.8\,T where the magnetic order is suppressed \cite{gass2020field, wolter2017field, bachus2021AngledependentThermodynamics}.

%The magnetic transition retains  its (weak) first-order character also in fields
%up to at least 5T (cf. Fig. 2b). It raises question about the influence of this (weak) first-order character on the field induced magnetic transition at 0Hc1
%= 7.8T, where the magnetic order is suppressed. Indeed this transition has previously been interpreted as a quantum critical point assuming a second order phase transition [13, 39

\subsection{Structural phase transition at $T_d$} At higher pressures $p \geq$ 95\,MPa the system undergoes a structural phase transition which is accompanied by a strong dimerization of Ru-Ru bonds \cite{bastien2018pressure, biesner2018detuning} and a collapse of the magnetic susceptibility. The transition at $T_d$ is strongly first order as reflected by the large and discontinuous changes in $\chi$ and the pronounced thermal hysteresis. The strongly first-order character of $T_d$ also provides a rational for the existence of a finite pressure range 95\,MPa $\leq p \leq$ 140\,MPa of an inhomogeneous coexistence, below $T_d$, of a dimerized non-magnetic phase and a non-dimerized phase. The latter phase shows the same signatures of magnetic order as revealed for lower pressures $p <$ 95\,MPa, albeit significantly reduced in size and a $T_N$ which hardly changes with pressure. In the range of phase coexistence, the relative volume fraction of the dimerized phase vs. the non-dimerized magnetic phase grows with increasing pressure.\\

\subsection{Healing effect of pressure-temperature treatment on a multiple-$T_N$ state}

 The occurrence of multiple magnetic transitions in $\alpha$-RuCl$_3$ has been assigned to stacking faults, i.e., different stacking patterns of the honeycomb layers that coexist in separated domains of the material \cite{cao2016low}. The results of the magnetic susceptibility on crystal \#2 (cf.\,Fig.\,5(a,b)), yielding two transitions, as opposed to a single transition for crystal \#1, indicate a distinctly higher concentration of these stacking faults for crystal \#2. By applying moderate pressure at low temperatures (20\,K), the relative size of the anomaly in $\chi$ at $T_{N1}$ was found to grow at the cost of the anomaly at $T_{N2}$, indicating a change in the relative volume fraction of the corresponding magnetic phases. This 'transformation process' towards a single-$T_N$ state continues with increasing pressure and is almost completed at $p$ = 95\,MPa. At this pressure there is only a small signal related to the transition at $T_{N2}$ visible. The pressure-induced single-$T_N$ magnetic state remains stable as long as the crystal is kept at low enough temperatures -- in this experiment below 20\,K -- but adopts its original multi-$T_N$ state after being warmed up to 300\,K (at ambient pressure). A full transformation into a truly 'single-$T_N$' state at $p$ = 0 can be obtained, however, by a temporary application of a pressure of 140\,MPa at a temperature of 50\,K. These observations indicate a small energy barrier (of order 10\,meV), associated with the stacking faults, separating a meta-stable single-$T_N$ state from the distorted multi-$T_N$ state. This conclusion is consistent with time-dependent studies (not shown) of the susceptibility on crystal \#2 after releasing the pressure at $T$ = 20\,K from 95.0\,MPa (or lower) to ambient pressure resulting in relaxational behavior with a time constant of several hours.

\section{Conclusions} %\textendash\ ............................\\

Measurements of the magnetic susceptibility under hydrostatic (He-gas) pressure conditions reveal a rapid suppression of the magnetic ordering
temperature in $\alpha$-RuCl$_3$ with pressure from $T_N$ = 7.3\,K at $p$ = 0 to about 6.1\,K at $p \leq$ 94\,MPa. Importantly, however, the experiments show that $T_N$ cannot be completely suppressed upon further increasing the pressure due to the occurrence of the pressure-induced dimerization transition at $p \geq$ 104\,MPa. Furthermore, it is found that the magnetic susceptibility in the paramagnetic regime, and before dimerization, considerably increases with pressure. Based on our model calculations, we assign this behavior to a pressure-induced strengthening of the nearest-neighbor Heisenberg $|J|$ and off-diagonal anisotropic coupling $\Gamma$ together with a simultaneous weakening of the Kitaev $|K|$ and anisotropic $|\Gamma|$' couplings. These results
 indicate a region of competition between two long-range orders, AFM zigzag and FM, before entering the dimerized phase.

Thorough investigations of the magnetic transition at $T_N$ at varying pressures and magnetic fields reveal clear indications for a weak first-order transition. We interpret this experimental finding as an indication of a strong magnetoelastic coupling in $\alpha$-RuCl$_3$,
as also suggested by our calculations.
 Moreover, by comparative studies on a second single crystal, showing two instead of one magnetic phase transition, a single-$T_N$ state could be prepared by the application of certain pressure-temperature protocols. These observations indicate the important role stacking faults play for the type of long-range magnetic order that develops in $\alpha$-RuCl$_3$.

%----------------------%
%  8. Acknowledgement  %
%----------------------%
\section*{Acknowledgments}
This work was supported by the Deutsche Forschungsgemeinschaft (DFG, German Research Foundation) for funding through TRR 288 - 422213477 (project A05 and A06), Collaborative Research Center SFB 1143 (project-id 247310070) and W\"urzburg-Dresden Cluster of Excellence on Complexity and Topology in Quantum Matter - \textit{ct.qmat} (EXC 2147, project-id 390858490)
G.B. acknowledges financial support from the European Union’s Horizon 2020 research and innovation program under the Marie Skłodowska-Curie
Grant Agreement No. 796048.
We are very grateful to Mr. S. Seker (TU Dresden) for his contributions to synthesis, and to MSc. A. Brunner (TU Dresden) for the know-how preparation of chlorine capillaries.

%\newpage

\appendix
\begin{appendix}

\section{Sample preparation}

Samples \#1 and \#2 were prepared from Ru metal powder and gaseous Cl$_2$ by various tempering regimes. For sample \#1, fine $\alpha$-RuCl$_3$ powder was pre-synthesized from the elements by annealing at $450^\circ\text{C}$, then it was sealed off in another evacuated quartz ampoule and subjected to heating up to $1000^\circ\text{C}$ at the rate $60^\circ\text{C/h}$, soaking at $1000^\circ\text{C}$ for 5\,hrs, slow controlled cooling down to $1000^\circ\text{C}$ at the rate of $3^\circ\text{C/h}$, annealed at $600^\circ\text{C}$ for 96\,hrs and finally quenched into cold water. During this treatment the powder has transformed into hexagonal-shaped, relatively thick (up to 1\,mm) platelets of $\alpha$-RuCl$_3$, as confirmed by EDX and powder X-ray diffraction of grounded crystals.
Sample \#2 was produced from the stoichiometric amounts of the elements  that were evacuated in a quartz ampoule and heated up to $750^\circ\text{C}$ at the rate $60^\circ\text{C/h}$ in a two-zone furnace. After annealing at $750^\circ\text{C}$ for 2\,days, the temperature of one zone was decreased to $650^\circ\text{C}$ in 2\,hrs, and the ampoule was kept in the temperature gradient 750 – $650^\circ\text{C}$ for 5\,more days before water-quenching. Elongated platelet-like black crystals of $\alpha$-RuCl$_3$ formed in the ampoule’s ‘cold’ end. Their chemical composition and the monoclinic crystal structure were confirmed by EDX and single-crystal XRD, respectively.

\section{Determination of $T_N$($p$) from susceptibility measurements}

\begin{figure}[t]
    %\centering
	\includegraphics[width=0.8\columnwidth]{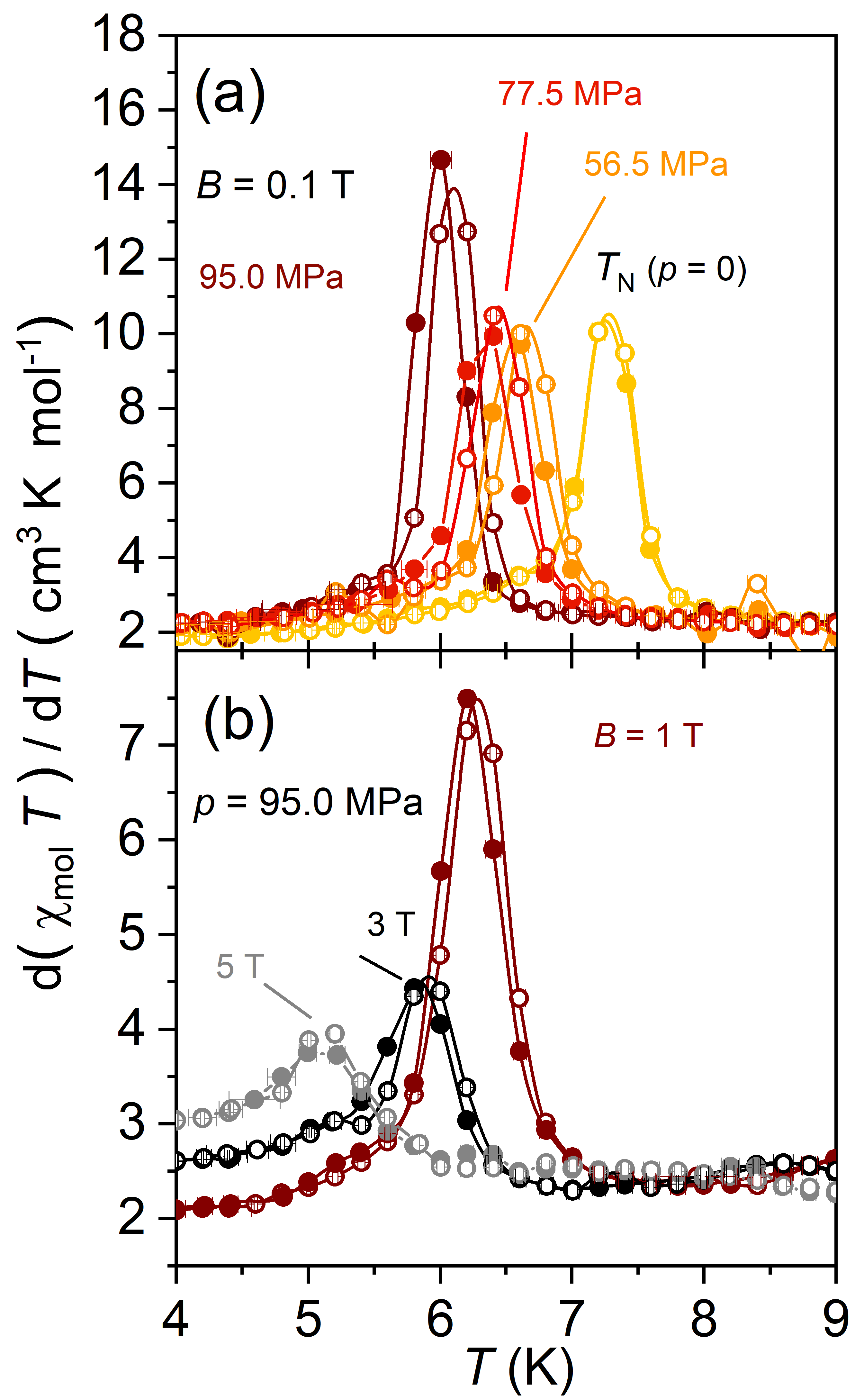}
	\caption{The temperature derivative of $\chi \cdot T$ of $\alpha$-RuCl$_3$ (crystal \#1). Upper part (a) shows data as a function of temperature at an in-plane field $B$ = 0.1\,T for various pressure values: $p$ = 0 (dark yellow circles), 56.5\,MPa (orange circles), 77.5\,MPa (red circles) and 95.0\,MPa (dark brown circles). The full (open) symbols represent the data taken with decreasing (increasing) temperature. Lower part (b) shows data as a function of temperature at $p$ = 95.0\,MPa for various magnetic fields applied parallel to the planes: $B$ = 1\,T (brown circles), 3\,T (black circles) and 5\,T (grey circles). The position of the maximum in d($\chi \cdot T$)/d$T$ is used to determine the transition temperature to antiferromagnetic order at $T_N$.}
	\label{App1}
\end{figure}

 Figure~\ref{App1} shows the low-temperature part of the magnetic susceptibility plotted as d($\chi \cdot T$)/d$T$ for crystal \#1. The data were taken at varying pressure values at an in-plane field $B$ = 0.1\,T (Fig.~\ref{App1}a) and at different in-plane magnetic fields of $B$ = 1, 3, and 5\,T at $p$ = 95.0\,MPa (Fig.~\ref{App1}b. At each pressure value data have been taken upon decreasing (full circles) and increasing temperature (open circles). By assigning the transition temperature to the maximum in d($\chi \cdot T$)/d$T$, we find for the heating runs $T_N$ = 7.33\,K ($p$ = 0), 6.65\,K (56.4\,MPa), 6.5\,K (77.5\,MPa) and 6.04\,K (95.0\,MPa). Consistent with the lack of hysteresis in the $\chi$ vs.\,$T$ data at $p$ = 0 shown in Fig.\,2, there is also no indication for hysteretic behavior by plotting these data as d($\chi \cdot T$)/d$T$ vs.\,$T$ in Fig.~\ref{App1}a. In contrast, the phase transition anomaly revealed at finite pressures of 56.5, 77.5, and 95\,MPa all show a small but distinct hysteresis upon cooling and warming. Hysteretic behavior is revealed also in the data taken in fields of $B$ = 1\,T, 3\,T and 5\,T at $p$ = 95.0\,MPa, cf.\,Fig.~\ref{App1}b.

\section{Signatures of the structural transition at $T_s$ in thermal expansion}

\begin{figure}[t]
    %\centering
	\includegraphics[width=0.8\columnwidth]{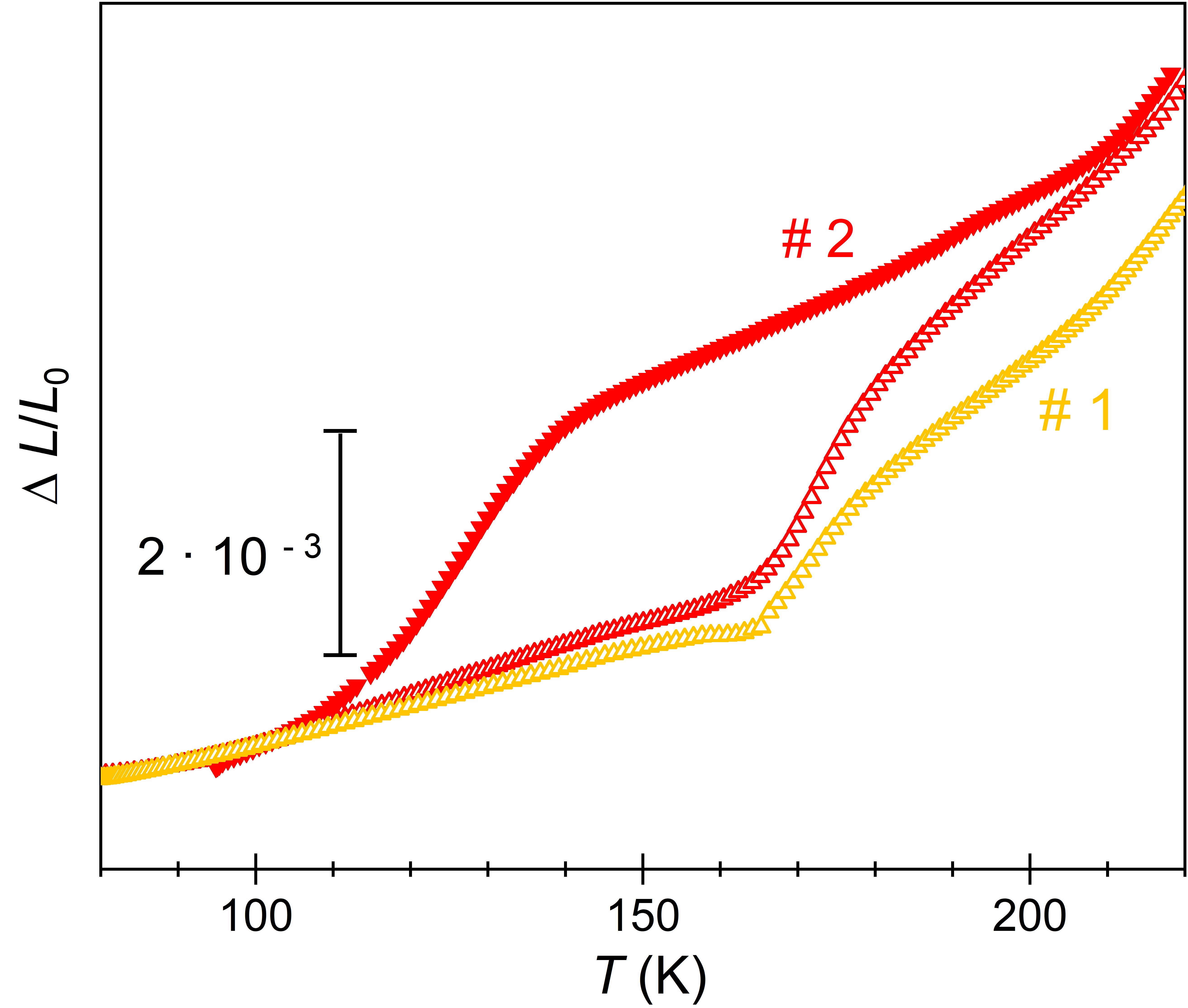}
	\caption{Thermal expansion $\Delta L(T) / L_0$ = ($L$($T$) - $L$(300\,K))/$L$(300\,K) for crystal \#1 (yellow triangles) and \#2 (red triangles) of $\alpha$-RuCl$_3$. The full (open) symbols represent the data taken with decreasing (increasing) temperature.}
	\label{App2}
\end{figure}

 Figure~\ref{App2} shows the relative length changes $\Delta L$($T$)/$L$ for crystal \#2 (red triangles) measured as a function of temperature for 80\,K $\leq$ $T$ $\leq$ 220\,K. The data reveal slightly broadened step-like changes of order $2\cdot 10^{-3}$ which are accompanied by a pronounced thermal hysteresis, indicating a first-order phase transition. The data are consistent with results reported in literature \cite{Widmann19, He18, gass2020field}. On increasing the temperature (open triangles) the midpoint of the transition is around 172\,K whereas it is around 126\,K on decreasing temperature (closed triangles) for \#2. A jump of similar size is observed also for crystal \#1 upon warming (yellow triangles), where it coincides with the transition observed in the magnetic susceptibility at $T_s$, cf.\,Fig.\,1.\\

\end{appendix}

\bibliography{bibly}

\end{document}